# Coherent control of superradiance from nitrogen ions pumped with femtosecond pulses

An Zhang,[1] Qingqing Liang,[2] Mingwei Lei,[1] Luqi Yuan,[3] Yi Liu,[2, #] Zhengquan Fan,[2] Xiang Zhang,[2] Songlin Zhuang,[2] Chengyin Wu[1,4], Qihuang Gong[1,4], Hongbing Jiang[1,4, *]

[1]State Key Laboratory for Mesoscopic Physics, School of Physics, Peking University, Beijing 100871, China
[2] Shanghai Key Lab of Modern Optical System, University of Shanghai for Science and Technology, 516, Jungong Road, 200093 Shanghai, China
[3]Ginzton Laboratory, Stanford University, Stanford, California 94305, USA
[4]Collaborative Innovation Center of Extreme Optics, Shanxi University, Taiyuan, Shanxi 030006, China
[#] yi.liu@usst.edu.cn, *hbjiang@pku.edu.cn

**Abstract**

Singly ionized nitrogen molecules in ambient air pumped by near-infrared femtosecond laser give rise to superradiant emission. Here we demonstrate coherent control of this superradiance by injecting a pair of resonant seeding pulses inside the nitrogen gas plasma. Strong modulation of the 391.4 nm superradiance with a period of 1.3 fs is observed when the delay between the two seeding pulses are finely tuned, pinpointing the essential role of macroscopic coherence in this lasing process. Based on this time-resolved method, the complex temporal evolution of the macroscopic coherence between two involved energy levels has been experimentally revealed, which is found to last for around 10 picoseconds in the low gas pressure range. These observations provide a new level of control on the "air lasing" based on nitrogen ions, which can find potential applications in optical remote sensing.

Cavity-free lasing of ambient air (nitrogen or oxygen) pumped by ultrafast laser pulses has opened the perspective to perform optical remote sensing with a directional coherent laser beam emitted from the sky to the ground observer [1-12]. This is in strong contrast with the traditional Light Detection and Ranging (LIDAR), where the incoherent scattered photons or the fluorescence of the target molecules in the atmosphere excited by a forward-propagating laser was collected by a telescope on the ground. The directionality and coherent nature of the backward propagating air laser beam promises significant improvement of the sensitivity of remote sensing [1-4, 13]. Up to now, both oxygen and nitrogen have been shown to be able to emit lasing radiation under proper pump conditions. The bidirectional near-infrared emission at 845 nm of oxygen atoms has been identified due to population inversion promoted by two-photon dissociation followed by a resonant two photon absorption of the oxygen molecules [1, 2, 6, 14]. In the meantime, the lasing emission of singly ionized nitrogen molecules has been intensely debated concerning its nature and mechanism [3, 7-9, 15-20]. This is partially due to the complexity of this emission process, where the electronic, vibrational, and rotational freedom of the nitrogen molecules are all coupled by the strong laser field [3, 15, 16, 19]. Moreover, it is found that the pump laser wavelength also plays a critical role and different mechanisms can be involved for 800 nm or mid-infrared pump pulses [16, 18].

A remarkable observation of the nitrogen ion lasing is that a delayed seeding pulse around 391.4 nm can be significantly amplified, up to a few hundred times, as to its energy [5, 15-17]. Several interpretations have been proposed for this optical amplification, such as population inversion mediated by the intermediate $A^2\Pi_u$ state [8, 16], transient inversion due to molecular rotational wavepackets [21, 22], inversion of partial rotational quantum levels without entire inversion of the $B^2\Sigma_u^+$ and $X^2\Sigma_g^+$ electronic levels [19]. Recently, it was revealed with time-resolved technique that the amplified 391.4 nm radiation is actually largely lags behind the injected seeding pulse [5, 17, 23]. More specially, the amplified pulse shows a pressure dependent temporal profile. With increasing nitrogen gas pressure $p$, both the built-up time $\tau_d$ and the pulse width $\tau_w$ decrease as $p^{-1}$, while the emission peak intensity scales up as $p^2$, which are characteristic for superradiance [17, 23-25]. Superradiance refers to a cooperative emission process of an ensemble of emitters where a large macroscopic coherence built up, which is normally initialized by spontaneous photons or externally seeding pulse [24, 25]. Macroscopic coherence between the two involved energy levels is characteristic for the superradiant radiation process. However, it has not yet been observed directly in the nitrogen-ions lasing experiments.

In this paper, we applied coherent control scheme to the nitrogen ions lasing system by injecting a sequence of seeding pulses at the resonance wavelength (391.4 nm) after the 800 nm pump pulse. In the experiment, it was found that the amplified 391.4nm emission experiences strong intensity modulation with a period of 1.3 fs as the delay between the two seeding pulses were finely tuned. This period corresponds to the transition frequency between the $B^2\Sigma_u^+$ and $X^2\Sigma_g^+$ electronic levels. A long-lived coherence between the two electronic states was revealed since the intensity modulation was observed up to a delay of ~ 10 ps in low gas pressure. We theoretically considered the nitrogen ions as a two-level system, interacting resonantly with the two seeding pulses. It is shown that the relative phase (delay) of the two seeding pulse causes a dynamic modulation on the coherence, which leads to the intensity variation during the superradiant emission process. Based on the modulation contrast, we measured the built-up and decay of the coherence between the $B^2\Sigma_u^+$ and $X^2\Sigma_g^+$ electronic levels in the nitrogen ions.

The laser pulses are delivered by a Ti: Sapphire amplification system at 800 nm wavelength with pulse energy of 3.6 mJ, pulse duration of 40 fs and repetition rate of 1 kHz. The output laser beam is split into two parts with a beam splitter. The main pulse of 2 mJ energy is used as the pump pulse to ionize nitrogen gas. Another beam passed through a 0.1 mm thick Beta barium borate (BBO) crystal to generate second harmonic pulse around 391 nm. The second harmonic pulse is further split into two beams. Each UV beam pass through a high precision mechanic delay line with a resolution of 10 nm, corresponding to a temporal resolution of 33 attoseconds. Then the two UV beams were combined together by another beam splitter. The pump pulse at 800 nm and the seeding pulse sequence were combined with a dichromatic mirror, presented in Fig. 1. The relative time delay between the three pulses can be changed with the two mechanic delay lines. The three pulses were focused together with an $f = 200$ mm lens into a gas chamber filled with nitrogen gas at different pressures. A visible plasma was formed due to photoionization by the intense laser field. After the nonlinear interaction inside the gas chamber, the radiation around 391.4 nm was analyzed with a spectrometer. Since we concentrated on UV emission from the nitrogen ions, the strong pump pulse around 800 nm and the accompanying white light were suppressed with proper glass filters before detection.

We first characterized the optical amplification and superradiance behavior of the 391.4 nm emission by injection of one seeding pulse. The gas pressure is 7.2 mbar and the delay time of the seeding pulse to the pump pulse $\tau_{ps}$ is set fixed about 0.75 ps. The results are presented in Fig. 2. In the presence of the seeding pulse, a strong amplified emission around 391.4 nm and 388.5 nm was obtained, which correspond to the P and R branch of the $B^2\Sigma_u^+$ to $X^2\Sigma_g^+$ transition (Fig. 2(a)). Time-resolved measurement of this amplified 391.4 nm was performed with a crossed correlation method similar to ref [5, 17], where the 391.4 nm pulse was frequency mixed with another weak 800 nm probe pulse in a sum-frequency BBO crystal and the sum-frequency generation signal at 263 nm was measured for varying delay between the 391.4 nm radiation and the weak 800 nm probe pulse. We presented the results in Fig. 2(b) for different intensities of the seeding pulse. We note that the amplified 391.4 nm emission lags largely behind the femtosecond seeding pulse, in agreement with the previous reports [17, 23]. With increased intensity of the seeding pulse, it was observed that the emission built up more rapidly. This is expected since the seeding pulse serves as the initial trigger for the formation of the macroscopic coherence during the superradiance process [25, 26]. The sharp emission peaks around 8.4, 12.6, and 16.8ps were revealed in our measurement, which corresponds to the quantum alignment of the nitrogen molecular ions [21, 27].

Then we injected a pair of seeding pulses inside the nitrogen gas plasma. The spectrum of the emission was recorded as a function of the time delay $\tau_{ss}$ between the two seeding pulses. In this experiment, the delay of the first seeding with respect to the 800 nm pump was fixed to be around 0.3 ps. We present in Fig. 3 the experimental results. In Fig. 3(a), strong variation with modulation depth about 48% of both the *P* and *R* branch is observed when the delay was finely changed, with a delay $\tau_{ss}$ in the vicinity of 3 ps. The modulation period was found to be 1.3 fs, in consistent with the oscillation period of 391.4 nm polarization. We have performed such measurements at different delays and some results are shown in Fig. 3 (b)-(c). Significant modulation with 1.3 fs period can be observed for longer delay $\tau_{ss}$ up to 13 ps in our experiments, demonstrating coherent control of this superradiance phenomenon.

For the purpose of understanding the fundamental mechanism of this coherent control effect, here we consider a simple model of two-level system, where level 1 is the ground state $X^2\Sigma_g^+$ and level 2 is the excited state $B^2\Sigma_u^+$. We assume that the system has initial population distribution $\rho_{11}(0)$ and $\rho_{22}(0)$ as well as the initial coherence $\rho_{12}(0)$ at *t* = 0 after the first 800 nm pump field passes through the medium. We consider the process that the first seeding pulse $E_1(t)e^{i\omega t} + c.c.$ interacts with this two-level system. For simplicity, we assume that the center frequency of the seeding field is resonant with the two-level transition, i.e. $\omega = \omega_{12}$. Therefore, by defining $\rho_{12} = \sigma_{12}e^{i\omega t}$, one can write the evolution equation under the rotating-wave approximation:

$$\frac{\partial \sigma_{12}}{\partial t} = -i\frac{\wp E_1}{\hbar}(\rho_{11} - \rho_{22}), \qquad (1)$$

where $\wp$ is the dipole moment.

The seeding field is ultrashort with a temporal width Δ (from center to nearly zero). It locates at time $\tau_1$ and the electric field is described as $E_1(t - \tau_1)e^{i\omega(t-\tau_1)} + c.c$. In the limit of $\wp E_1\Delta/\hbar \ll 1$, one can integrate Eq. (1) with the first-order approximation, and therefore obtain the expression of the coherence after the passage of the first seeding field passes (t = $\tau_1$ + Δ):

$$\sigma_{12}(\tau_1 + \Delta) \approx -i[\rho_{11}(\tau_1) - \rho_{22}(\tau_1)]\frac{\wp}{\hbar}\int_{\tau_1-\Delta}^{\tau_1+\Delta} E_1(t')dt' e^{-i\omega\tau_1} \equiv \rho_{12}^I(-ie^{-i\omega\tau_1}). \qquad (2)$$

Here $\rho_{12}^I$ gives the amplitude of the coherence built after the first seeding field leaves the medium. The amplitude of the coherence at the moment that the seeding passes is dependent on the population distribution of the medium, the seeding field intensity, and the seeding field profile. Once the coherence is prepared by the first seeding pulse, it starts to grow up due to the optical gain $g(t)$ in the system, and the phase of the coherence is oscillating at the frequency $\omega$, i.e. $\rho_{12}(t) = -i\rho_{12}^I e^{i\omega(t-\tau_1)}e^{\int_{\tau_1}^{t} dt' g(t')}$. Before the second seeding pulse comes at $t = \tau_2$, we assume that its amplitude becomes $\rho_{12}^{II}$ ($|\rho_{12}^{II}| > |\rho_{12}^I|$ due to the optical gain of the nitrogen ion system). Next, we consider the second seeding field $E_2(t - \tau_2)e^{i\omega(t-\tau_2)} + c.c.$ coming at $t = \tau_2$ with the same pulse width Δ. The delay between two seed fields is much larger than the pulse width, i.e. $\tau_2 - \tau_1 \gg \Delta$. We again assume that $\wp E_2\Delta/\hbar \ll 1$, so one can integrate Eq. (1) and find the coherence after the passage of the second seeding field:

$$\sigma_{12}(\tau_2 + \Delta) \approx -i\rho_{12}^{II} e^{-i\omega\tau_1} - i[\rho_{11}(\tau_2) - \rho_{22}(\tau_2)]\frac{\wp}{\hbar}\int_{\tau_2-\Delta}^{\tau_2+\Delta} E_2(t')dt' e^{-i\omega\tau_2}$$
$$\equiv -i\rho_{12}^{II} e^{-i\omega\tau_1} - i\rho_{12}^{III} e^{-i\omega\tau_2}. \qquad (3)$$

Here $\rho_{12}^{III}$ represents the net amplitude change of the coherence induced by the injection of the second seeding pulse. The emission from the coherence is collected by the detector and the signal is measured after integration. The signal, which integrates the intensity of the emission over time, is $I = \int_0^\infty |E_{emission}|^2 dt \propto \int_0^\infty |\rho_{12}(t)|^2 dt$. As for the illustration, we assume that the emission lasts much longer than the time scale of $\tau_{1,2}$, so the signal is mainly contributed from the coherence built up after the second seeding pulse passes. Without loss of generality, we can write the signal from the coherence in Eq. (3)

$$I \propto |\rho_{12}^{II}|^2 + |\rho_{12}^{III}|^2 + (\rho_{12}^{II}\rho_{12}^{III*}e^{i\omega\tau_{ss}} + c.c) \qquad (4)$$

where $\tau_{ss} = \tau_2 - \tau_1$ is the time delay between two probe fields. One can see that the signal is beating at the frequency ω versus $\tau_{ss}$. It is now clear that the strong 1.3 fs modulation of the superradiance (Fig. 3) origins from the coherent interaction of the second seeding pulse with the macroscopic polarization formed by the first seeding pulse, where their relative phase is determined by the delay $\tau_{ss}$.

The modulation contrast of Fig. 3 can reflect the evolution of the macroscopic polarization in the nitrogen ions after the first seeding pulse. We therefore measured this contrast as a function of the seeding pulse time delay $\tau_{ss}$. We defined the modulation contrast η as the absolute value of $\eta = (I_{max}-I_{min})/(I_{max}+I_{min})$, where $I_{max}$ and $I_{min}$ are maximum and minimum 391.4 nm intensity. From Eq. (4), it gives $\eta = 4|\rho_{12}^{II}||\rho_{12}^{III}|/\left[|\rho_{12}^{II}|^2 + |\rho_{12}^{III}|^2\right]$. One therefore can see that the modulation contrast is supposed to be the maximum once $|\rho_{12}^{II}| = |\rho_{12}^{III}|$. In Fig. 4, we presented the results for three different intensity radio between the two seeding pulses. For the case of $I_{s2} = I_{s1}$, a monotonous decrease of the modulation contrast was observed. For larger intensity of the second seeding pulse of $I_{s2} = 2I_{s1}$ and $I_{s2} = 4I_{s1}$, we noticed that the contrast first increases and then decreases. A full understanding of these results will only be possible in the framework of a complete Maxwell-Bloch equation describing the interaction of the resonant seeding fields with the two-level molecular system, as well as the formation process of the superradiance [25, 26]. Here we would like to discuss qualitatively the main reasons underlying these observations. We first concentrate on the case where the two seeding pulses had the same energy. After the injection of the first seeding pulse, a seed-induced initial polarization $\rho_{12}^{I}$ is excited in the nitrogen plasma, with its amplitude depending on the electric field $\varepsilon_l$ of the seeding pulse and the population difference between the two levels $\rho_{11}(\tau_1)$-$\rho_{22}(\tau_1)$. In the presence of optical gain in the system, the seed-induced polarization evolves in the time domain, first grows to a maximum then decreases, with its maximum amplitude and temporal profile depending on the competition of gain, decoherence, and emission of superradiance [25, 26]. At time $\tau_2$, the second seeding was injected in the system. It interacts with the evolving two-level system and provokes an instantaneous polarization change $\rho_{12}^{III}$ inside the system, presented by the last term in Eq. 3. This net polarization change adds up coherently with the polarization under evolvement from $\tau_1$ to $\tau_2$, with their relative phase determined by ω($\tau_1$ - $\tau_2$). Strong modulation contrast of the final superradiance intensity is expected when $\rho_{12}^{III}$ has a comparable amplitude with the instantaneous polarization before the injection of the second seeding pulse $\rho_{12}^{II}$. In the case of $I_{s2} = I_{s1}$, we expect that $\rho_{12}^{III}$ is normally less than $\rho_{12}^{II}$ due to two reasons. First, the polarization after the first seeding pulse is increasing with time and we expect $\rho_{12}^{II}$ is much larger than the initial polarization $\rho_{12}^{I}$. At the same time, the second seeding pulse sees a decreased population difference $\rho_{11}(\tau_2)$-$\rho_{22}(\tau_2)$ compared to that at time $\tau_1$ and we expect $\rho_{12}^{III}$ is less than $\rho_{12}^{I}$. As a result, the modulation contrast decreases when the time delay $\tau_{ss}$ is increased, presented in Fig. 4. For increased intensity of the second seeding pulse, the polarization change $\rho_{12}^{III}$ due to its injection can be larger. Therefore, the amplitude of $\rho_{12}^{III}$ can be comparable to the amplified polarization $\rho_{12}^{II}$ due to the gain at a proper time delay $\tau_{ss}$, and hence a maximum modulation contrast can be expected. Obviously, for further increased intensity of the second seeding pulse $I_{s2} = 4I_{s1}$, the maximum contrast occurs for longer time delay. Therefore, the evolution of the macroscopic coherence is encoded in the variation of the modulation contrast presented in Fig. 4, while a quantitative understanding will be possible with a complete modeling and numerical simulation.

In conclusion, we demonstrated coherent control of the nitrogen ions "air laser" with a pair of seeding pulses at the resonant wavelength of the $B\,^2\Sigma_u^+$ to $X\,^2\Sigma_g^+$ transition. The superradiant emission at 391.4 nm shows a strong intensity modulation with a period determined by the transition frequency when the relative delay (phase) between the two seeding pulses is varied. With this method, the built-up and decay of the macroscopic coherence between two electronic levels have been experimentally revealed and it has been found to remain for tens of picosecond. This work highlights the essential role of the macroscopic coherence in the nitrogen ions lasing process and provides a new level of control on the nitrogen ions "air laser", which can be beneficial for its application in remote sensing.

**Acknowledgement**

The work is supported in part by the National Natural Science Foundation of China (Grants No. 11574213, No. 41527807, and No. 61590933), Innovation Program of Shanghai Municipal Education Commission (Grant No. 2017-01-07-00-07-E00007), and Shanghai Municipal Science and Technology Commission (No. 17060502500). Y. Liu acknowledges the support by The Program for Professor of Special Appointment (Eastern Scholar) at Shanghai Institutions of Higher Learning (No. TP2014046). The authors acknowledge stimulating discussion with Prof. Andre Mysyrowicz of Ecole Nationale de Technology Avancee in France.

An Zhang and Qingqing Liang contributed equally to this work.

**Figure 1**

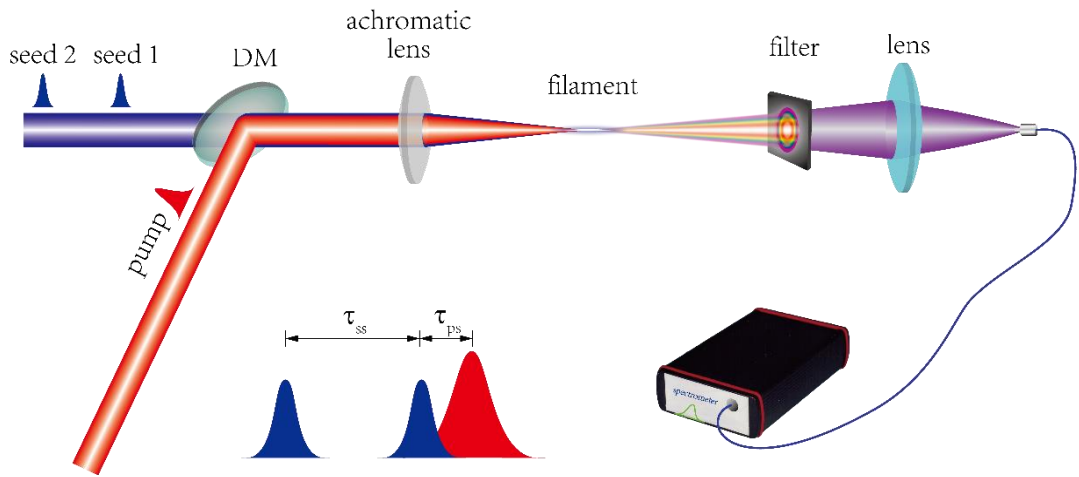

Fig. 1. Experimental setup for coherent control of the nitrogen ions superradiance. The pump pulse at 800 nm and the seeding pulses pair around 391 nm were combined together by a dichromatic mirror (DM). They were focused by an achromatic lens of $f$ = 200 mm in a gas cell filled with nitrogen gas. The forward radiation was spectrally filtered and further measured with a spectrometer. The relative delay between the pump and the first seeding pulse $\tau_{ps}$ and the delay between the two seeding pulse $\tau_{ss}$ can be separately controlled with two mechanic delay lines.

**Figure 2**

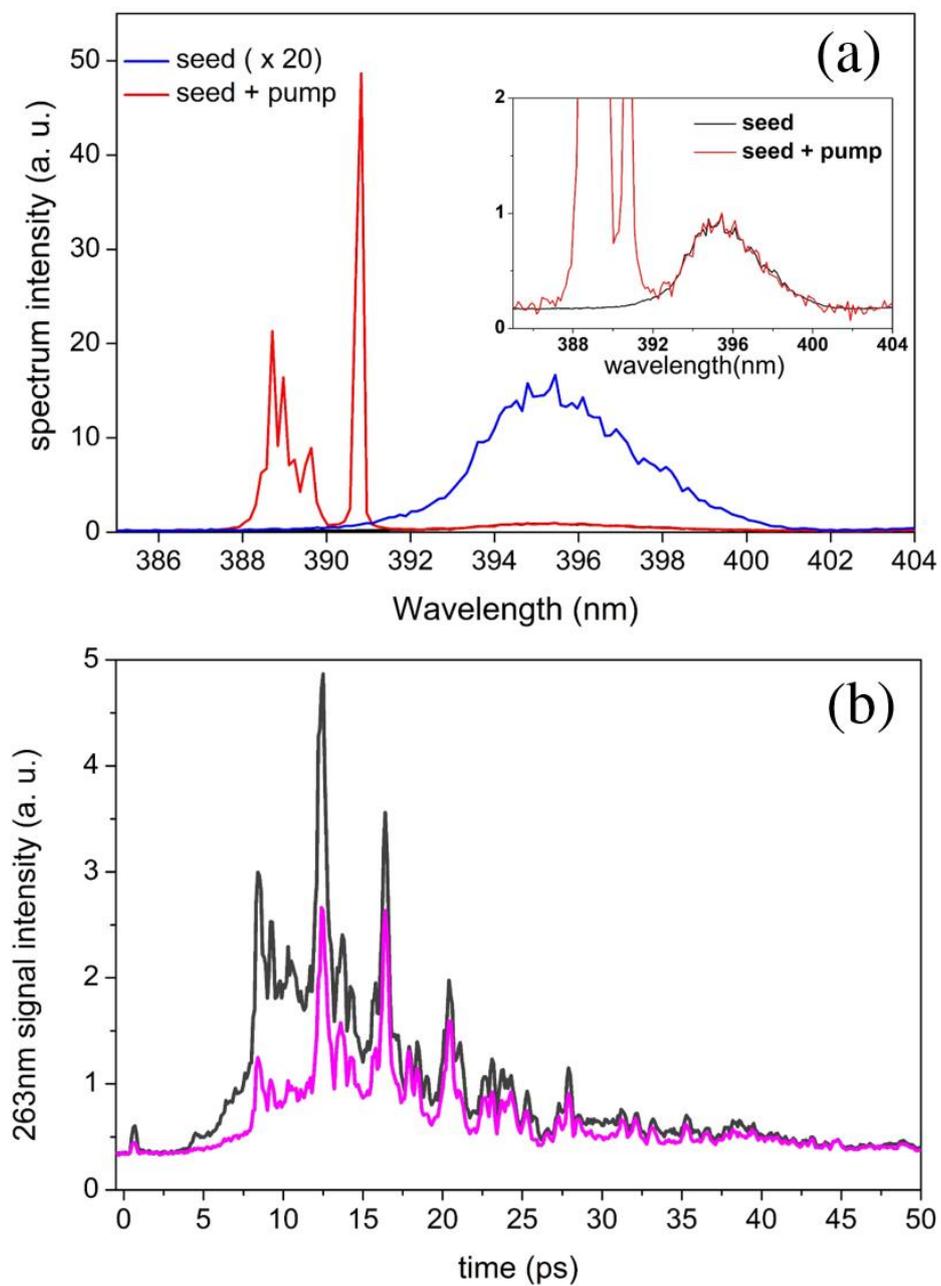

Fig. 2. (a) Amplification of the seeding pulse. The intensity of the seeding pulse was multiplied by a factor of 20 for easy comparison. Inset, comparison of seeding pulse and the amplified emission with zoomed vertical scale (b) Time-resolved 391.4 nm signal measured by cross-correlation method with two different seeding intensity. The seeding pulse intensity was doubled from the pink curve to the black curve. The weak peak around 0.75 ps corresponds to the seeding pulse.

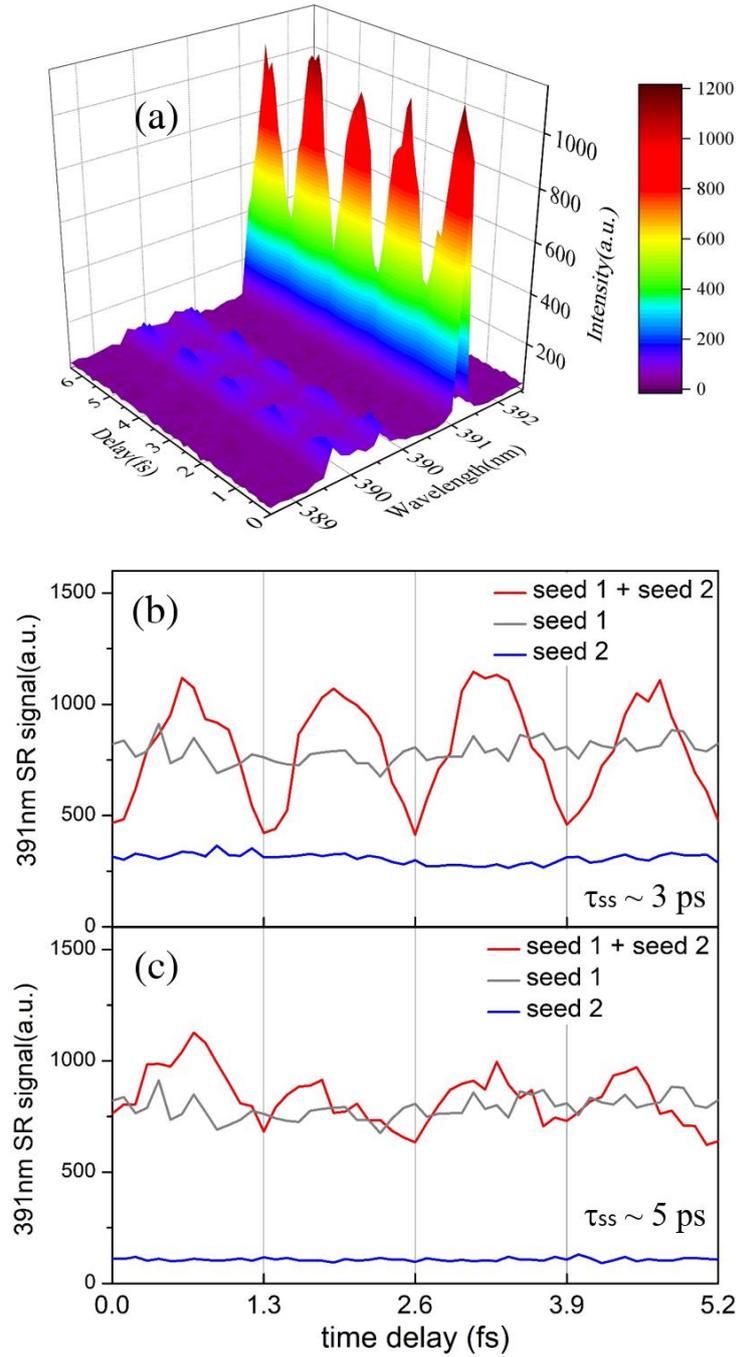

Fig. 3. (a) Spectrum of the forward UV emission around 391.4 nm as a function of the delay $\tau_{ss}$ between the two seeding pulses. The nitrogen gas pressure was 4 mbar. (b) and (c), Spectrum intensity at 391.4 nm versus fine tuning of $\tau_{ss}$ for different rough time delay $\tau_{ss}$. In (a) and (b), $\tau_{ss}$ is about 3 ps while it is about 5 ps for (c).

**Figure 4**

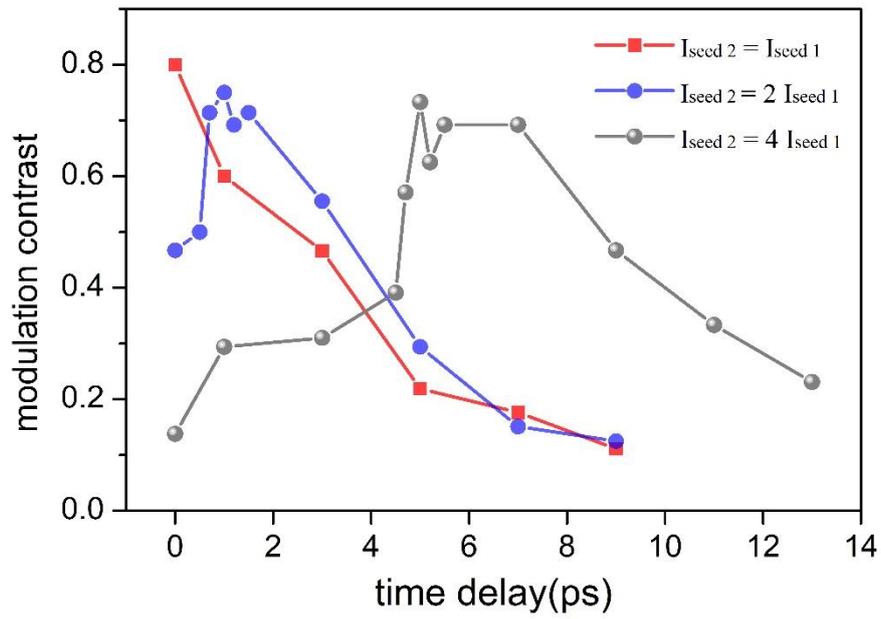

Fig. 4. Experimental result of the superradiance modulation contrast as a function of the time delay $\tau_{ss}$, for different ratio of intensity between the two seeding pulses. The points are experimental data, and the lines are for eye guide.